\documentclass[twocolumn,aps,prb,showpacs]{revtex4}

\usepackage{graphicx}
\usepackage{amssymb}
\usepackage{amsmath}
\usepackage{color}

\font\mycal=cmsy10

\def\e{\varepsilon}

\def\G{\mbox{\mycal G}\,}

\begin{document}

\title{Persistent Current in a Ring Coupled to an External Fermionic Reservoir}

\author{Bernhard Wunsch}
\affiliation{I. Institute of Theoretical Physics, University
of Hamburg, Jungiusstr. 9, D--20355 Hamburg, Germany}

\author{Alexander Chudnovskiy}
\email{achudnov@physnet.uni-hamburg.de}
\affiliation{I. Institute of Theoretical Physics, University
of Hamburg, Jungiusstr. 9, D--20355 Hamburg, Germany}

\begin{abstract}  \vspace{5mm}  We study the energy spectrum and the 
  persistent current in an ideal one-dimensional mesoscopic ring
  coupled to a fermionic reservoir. We find that the tunnel coupling in
general leads to the suppression of the persistent current. However,
with increasing coupling, the effective level structure of the
ring coupled to the reservoir changes and quasistates with a sharp eigenenergy
develop.
Depending on the number of ring states coupled to the reservoir this
results in a nonzero persistent current even at very large tunneling
between the ring and the reservoir.
\end{abstract}


\maketitle

Experiments on mesoscopic rings enable to study quantum effects based
on phase coherence like the Aharonov-Bohm effect
\cite{Aharonov59:485,Holleitner01:256802} and persistent currents
\cite{Lorke00:2223,current}.
As phase coherence is the precondition of these phenomena, the
influence of decoherence is of major interest.
Recently, the suppression of quantum coherence in a mesoscopic system due to its coupling
to an external macroscopic reservoir attracted much attention
\cite{Chudnovskiy}.
If a small mesoscopic system (quantum dot, quantum ring) is coupled
by tunneling to an external reservoir of fermions (a lead) a phenomenon
of level attraction is known to occur, which results in changes of occupation
numbers, statistics of energy levels, and eventually the
transport properties through the mesoscopic device.
In the present paper we investigate the effects of the level
attraction due to the tunnel coupling to an external reservoir on the
persistent current in a mesoscopic ring.

A ring coupled to a reservoir was investigated previously within the
scattering matrix approach \cite{Buttiker},
in which the ring is coupled via an ideal wire to the dissipative
reservoir. However, the development
of long living states for strong coupling was not discussed there.

\begin{figure}[h]
\begin{center}\leavevmode
\includegraphics[angle=0,scale=.25]{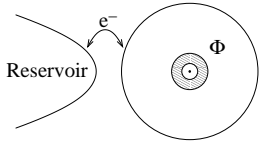}
\caption{Setup for a coupled ring described by the tunnel Hamiltonian formalism}
\label{fig:compButtikerTunnel}
\end{center}
\end{figure}

The setup studied in this work is shown in figure
\ref{fig:compButtikerTunnel}.  Since the main purpose of this paper is
to study the influence of decoherence introduced by a tunnel contact
to a fermionic reservoir (a lead) we confined ourselves to an
independent spinless electron model and assumed the ring to be
one-dimensional.
This simple model captures already
the main features of the energy spectrum and the persistent current measured
on rings in the ballistic transport regime\cite{Lorke00:2223}.
The Hamiltonian of our model has the following form:
{\small 
\begin{eqnarray}
\hat{H}&=& \sum_{m}\varepsilon_m \hat{a}_m^+
\hat{a}_m + \sum_{r} E_r \hat{b}_r^+ \hat{b}_r + \notag\\
& &+\sum_{m,r}(t_{m ,r} \hat{a}_m^+ \hat{b}_r +
h. c.)
\label{hamiltonian}
\end{eqnarray}
}
where $\hat{a}_m^+$ and $\hat{a}_m$ ($\hat{b}_r^+$ and $\hat{b}_r$)
are the creation and annihilation operators for electrons in the ring
(reservoir) with quantum number $m$ ($r$). The eigenfunctions of the
isolated ring are given by $\phi_m=e^{i\,m \varphi}$, where $m$ denotes
the angular momentum in the ring. The corresponding eigenenergies are
given by $\varepsilon_m=4 E_0
\left(m+\frac{\Phi}{\Phi_0}\right)^2$, where $\Phi$ denotes the magnetic flux through the
ring, $\Phi_0=\frac{h}{e}$ represents the magnetic flux
quantum. The energy scale is given by $\quad E_0=\frac{\hbar^2}{8
  m^* R^2}$. The flux dependence of the Hamiltonian
as well as the length of the ring is exclusively contained in the
eigenenergies of the ring.
The energies in the reservoir are denoted by $E_r$.

Following earlier discussion \cite{Shahbazyan} it can be shown that
for a small size of the contact between ring and reservoir, the
tunneling matrix elements depend very weakly on the reservoir
quantum number $r$. The dependence on the quantum number in the ring $m$
can be estimated as $t_{m,r}\propto\frac{2}{m}\sin{m \varphi_0}$,
where $\varphi_0$ describes the angular size of the tunnel contact.
In this paper we set the tunnel matrix
elements constant for a given range of angular momenta of the ring
eigenstates.  The tunnel matrix
elements for other ring states are set to zero.

We point out, that different ring states couple to the
same states in the reservoir. This effective interaction between ring
states will determine the behavior for strong coupling, as shown in
the following.

Within the described model, a Dyson equation for the
Green's function can be solved exactly for
arbitrary tunneling strength.  The retarded Green's function  determines the spectral
density of the ring states $S_m(E)=-\frac{1}{\pi} Im\left(\G^{ret}_m(E)\right)$, from which
the total density of states (DOS) inside the ring $\rho(E)$ is obtained:
\begin{eqnarray}
\rho(E)&=&\sum_m S_m(E)=\frac{-\kappa \frac{\partial \xi(E)}{\partial E}}
     {\pi \left(1+\kappa^2 \xi(E)^2\right)}, \label{rhogeneral}
\end{eqnarray}
with $\xi(E)=\sum_{m_1}\frac{1}{E-\e_{m_1}}$ and $\kappa=\pi \nu |t|^2$.
The obtained results are also valid for
the strong coupling regime, in which the energy scale given by the
tunneling is of the order of or larger than the interlevel spacing
between consecutive ring states.  To avoid superimposing effects on
the DOS in the ring due to the band structure of the reservoir we
choose a constant density of states in the reservoir $\nu(E)=\nu$.

In the simplest case of only one ring state that couples to the
reservoir, the DOS is given by a Lorentzian centered
around the eigenenergy $\e_0$ of the isolated state
$\rho(E)=\frac{\kappa}{\pi \left((E-\e_0)^2+ \kappa^2\right)}$ with
the width given by the coupling energy $\kappa$.

The effect of coupling to the reservoir on the persistent current in the ring is
investigated by calculating the current density. The latter is obtained by
summing over the contributions of all ring states.  As the current carried
by an occupied isolated ring state is $I_m=-\frac{\partial
  \varepsilon_m}{\partial \Phi}$, the current density has the form
\begin{eqnarray}
     j(E)&=&\sum_m -\frac{\partial \varepsilon_m}{\partial \Phi}S_m(E)=
     -\frac{\kappa \,\frac{\partial \xi(E)}{\partial \Phi}}
     {\pi(1+\kappa^2 \xi(E)^2)}. \label{jgeneral}
\end{eqnarray}
The total current is then given by $I=\int_{-\infty}^{\infty} f(E) j(E)dE$,
where $f(E)$ denotes the Fermi distribution.

Let us first assume, that the coupling between ring and reservoir is
restricted to the two energetically lowest ring states.  This can be
motivated by selective tunneling with respect to the angular momentum
of the ring states as discussed above. Furthermore for a magnetic flux
close to $\Phi=\frac{\Phi_0}{2}$ this assumption is also a good
approximation as long as the energy gap to the higher lying ring
states is larger than the coupling energy $\kappa$. However we will
not limit the coupling strength in the following discussion.

\begin{figure}
\begin{center}
\includegraphics[angle=0,scale=0.7]{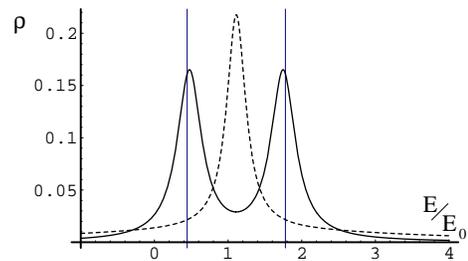}
\end{center}
\caption{DOS for two coupled states and fixed magnetic flux $\Phi=\frac{\Phi_0}{3}$
for weak coupling (full line, $\kappa=0.2\,E_0$ )
and strong coupling (dashed line, $\kappa=1.5\,E_0$).
The grid lines show the eigenergies in the isolated ring.}
\label{fig:2N_rho}
\end{figure}

If only two ring states couple to the reservoir, the system can be well
described by introducing two quasistates.
For weak coupling, their DOS is given by:
\begin{eqnarray}
S_{1/2}(E)&=&\frac{\kappa}{\pi\left(\left(E_{av} \pm \sqrt{\frac{(\Delta\e)^2}{4}-\kappa^2}
\right)^2+\kappa^2\right)}
\end{eqnarray}
with $
\Delta\e=\e_{m_1}-\e_{m_2};\quad E_{av}=E-\frac{\e_{m_1}+\e_{m_2}}{2}.
$
The strength of the coupling is characterized by the tunneling energy
$\kappa$ defined above, which has to be compared to
the interlevel spacing $\Delta\e$ of the coupled ring states.
Without coupling, the quasistates coincide with the eigenstates of the
isolated ring. The solid line in figure \ref{fig:2N_rho} shows the
DOS in the weak coupling regime defined by
$\frac{(\Delta\e)^2}{4}-\kappa^2>0$. In this regime the spectral
densities of the quasistates broaden with increasing coupling, thereby
approaching each other.

At the critical coupling $\kappa_{c}=|\frac{\Delta\e}{2}|$ the
spectral densities of the quasistates are equivalent. The dashed line
in figure \ref{fig:2N_rho} illustrates the DOS, if the coupling is increased
into the strong coupling regime with $\kappa>\kappa_c$.
One quasistate with a sharp eigenenergy
develops and is represented by the sharp peak in the DOS with a width
smaller than $\kappa$. The other quasistate contributes to the DOS
within a broad energy range of a width larger than $\kappa$.
The described phenomenon can be related to the Dicke effect in optics
\cite{Dicke}. The manifestation of the latter has been investigated recently
in context of resonant scattering and resonant tunneling in solid state systems
\cite{Shahbazyan}.
The behavior of the DOS is well described by the spectral densities
of the quasistates in the strong coupling regime: {\small
\begin{eqnarray}
S_{1/2}(E)=\frac{\kappa\mp\sqrt{\kappa^2-\frac{(\Delta\e)^2}{4}}}
{\pi\left(E_{av}^2+\left(\kappa \mp \sqrt{\kappa^2-\frac{(\Delta\e)^2}{4}} \right)^2\right)}
\quad.
\end{eqnarray}
}
It is important to notice that the energies of the isolated ring
states depend on the magnetic flux, while the coupling to the reservoir
is assumed to be independent of the magnetic flux. In particular, the
eigenenergies are degenerate at $\Phi=n \frac{\Phi_0}{2}$, so that by
changing the magnetic flux close to this degenerate value one finally
enters the regime of strong coupling, for any nonzero coupling strength.

\begin{figure}
\begin{center}
\includegraphics[angle=0,scale=0.7]{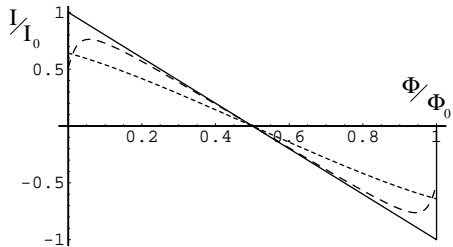}
\end{center}
\caption{Dependence of the persistent current on the magnetic
  flux for two coupled ring states. Parameters: $\mu=4\,E_0$; full
  line: $\kappa=0$; long dashed line: $\kappa=0.2\,E_0$; short dashed
  line: $\kappa=5\,E_0$.}
\label{fig:2N_I}
\end{figure}

The energy of the long living state depends on the magnetic flux, as
it is given by the average energy of the two coupled states.
Therefore, the system shows Aharonov-Bohm type behavior even in the
strong coupling regime.  Correspondingly the persistent current
saturates in the limit of strong coupling and does not vanish as
illustrated in figure \ref{fig:2N_I}.  In the limit of strong
coupling the long living state carries the current
$I=\frac{I_{m_1}+I_{m_2}}{2}$, while the broad state
carries a current of $I=\frac{I_{m_1}+I_{m_2}}{4}$. Therfore the value
of the saturated persistent current is either $\frac{1}{4}$ or
$\frac{3}{4}$ of the current in the isolated ring, depending on
whether the Fermi energy lies below or above the energy of the
long-living state.

A generalization of the simplified two level model is obtained by
considering the coupling of more ring states to the reservoir. Thereby at
least all states with an eigenenergy below the Fermi energy are
coupled to the reservoir.
In the strong coupling regime the system develops long living states
between nearest neighbors whenever the tunneling energy $\kappa$
exceeds the interlevel spacing between the corresponding eigenenergies
of the isolated ring. Like in the case of two coupled levels, the persistent
current saturates in the limit of strong coupling at a generally
nonzero value.  Thereby the saturation value of the persistent current
depends strongly on the number of coupled states.  It decreases with
increasing number of coupled ring states, but it also shows an odd-even
effect with the number of coupled states. Both features have
their origin in the alternating sign of the current carried by
consecutive ring states.

Now we consider the coupling of all ring states to the reservoir.
According to our previous discussion about the tunnel matrix
element this situation is realized in the limit of a
point contact.  It is an appealing feature of our model that an easy
analytical formula for the DOS in the ring and the current density can
be given for this limiting case:
\begin{eqnarray}
\rho(E)=\frac{\kappa}{\pi}
\frac{\frac{1}{x}\sin x \left(\cos\Phi-\cos\,x \right)+1-\cos\,x \cos\Phi}
{\kappa^2 \left(\sin\,x \right)^2 + \frac{4 E_0^2 x^2}{\pi^4}\left(\cos\Phi-\cos\,x \right)^2}
\label{rho}
\end{eqnarray}
with
$
x=\pi \sqrt{\frac{E}{E_0}}
$,
thereby we have set $\hbar=e=c=1$.
To obtain the DOS given in equation (\ref{rho}) we have used $\xi(E) = \sum_{m=-\infty}^{\infty}
\frac{1}{E-\e_m} =\frac{\pi^2}{2 E_0 x}
\frac{\sin x}{\cos \Phi -\cos x}$.

\begin{figure}
\begin{center}
\includegraphics[angle=0,scale=.7]{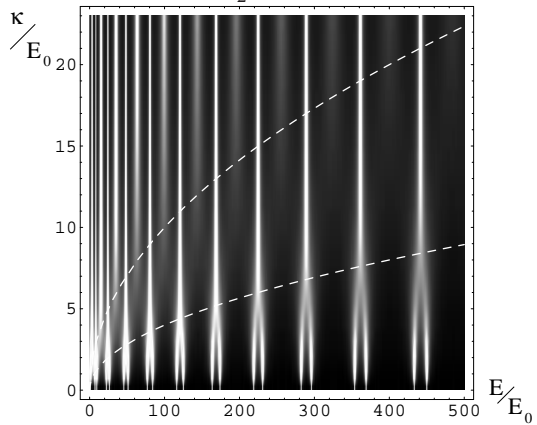}
\end{center}
\caption{Density plot of DOS as a function of the coupling for fixed
    magnetic flux $\Phi=0.4 \Phi_0$ (the magnitude of the DOS increases
  going from black to white).  For weak coupling the maxima are located at
  the eigenenergies of the isolated ring states whereas for strong coupling
  quasistates at $E=n^2\, E_0$ develop.
  The critical coupling thereby depends on the energy as $\sqrt{E}$
  with different prefactors for even or odd n as indicated by the
  dashed lines ($0.4 \sqrt{E}$, $\sqrt{E}$).}
\label{fig:ANrhoEkappa}
\end{figure}

For small coupling the DOS shows Lorentz broadened maxima around the
eigenenergies of the isolated ring states with a width given by the
coupling energy $\kappa$, in agreement with the coupling of two states.
Figure \ref{fig:ANrhoEkappa} illustrates the development of the DOS
with increasing coupling.  The maxima in the DOS belonging to
different ring states move together with increasing coupling and at a
critical coupling new maxima are formed at $E=(2m+1)^2 E_0$. Thereby the
critical coupling grows with energy proportional to $\sqrt{E}$, like
the interlevel spacing between the eigenenergies of the isolated ring.
At even stronger coupling additional quasistates are developed at
$E=(2m)^2 E_0$. The peaks in the DOS corresponding to these
quasistates are more and more pronounced with increasing coupling.
Their energies have also a special meaning in the energy spectra of
the isolated ring.  $E=(2n+1)^2\,E_0$ are the two fold degenerate
eigenenergies for $\Phi=\pm \frac{\Phi_0}{2}$, while $E=(2n)^2\,E_0$
are the two fold degenerate eigenenergies for $\Phi=n \Phi_0$. For
other values of the magnetic flux the energy spectrum is nondegenerate.

\begin{figure}
\begin{center}
\includegraphics[angle=0,scale=0.4]{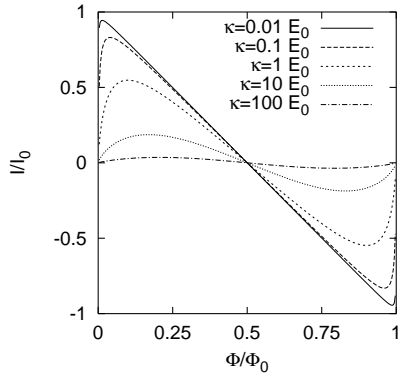}
\end{center}
\caption{Dependence of the persistent current on the magnetic flux
  for different coupling strengths and for coupling of all ring states.
  For all curves, there are five states below the Fermi energy $\mu=25 E_0$.}
\label{fig:ANIphi_article}
\end{figure}

In contrast to the coupling of a finite number of ring states, the
eigenenergies of the long living states in the strong coupling regime
are now independent of the magnetic field, which evidences for the
localization of those states in the ring.  Consequently, the
Aharonov-Bohm effect disappears, which is accompanied by a continuous
suppression of the persistent current with increasing coupling as
illustrated in figure \ref{fig:ANIphi_article}.

The authors are grateful to D. Pfannkuche for illuminating  discussions.
Financial support form SFB 508 is gratefully acknowledged.


\end{document}